\documentclass[twocolumn,american,english]{revtex4-1}
\usepackage[T1]{fontenc}
\usepackage[latin9]{inputenc}
\usepackage{textcomp}
\usepackage{amsmath}
\usepackage{graphicx}
\usepackage{colortbl}
\usepackage{natbib}
\makeatletter

\newcommand*\LyXZeroWidthSpace{\hspace{0pt}}
\providecommand{\tabularnewline}{\\}

\makeatother

\usepackage{babel}
\begin{document}

\title{Experimental Study of the Bottleneck in Fully Developed Turbulence}

\author{Christian Küchler$^{1}$}

\author{Gregory Bewley$^{2}$}

\author{Eberhard Bodenschatz$^{1,2, 3}$}

\affiliation{$^{1}$Max-Planck-Institute for Dynamics and Self-Organization, Göttingen,
Germany}
\affiliation{$^{2}$Cornell University, Ithaca, USA}
\affiliation{$^{3}$Georg August University Göttingen, Germany}

\begin{abstract}
The energy spectrum of incompressible turbulence is known to reveal
a pileup of energy at those high wavenumbers where viscous dissipation
begins to act. It is called the bottleneck effect \cite{donzis_bottleneck_2010,falkovich_bottleneck_1994,frisch_hyperviscosity_2008,kurien_cascade_2004,verma_energy_2007}.
Based on direct numerical simulations of the incompressible Navier-Stokes
equations, results from Donzis \& Sreenivasan \cite{donzis_bottleneck_2010}
pointed to a decrease of the strength of the bottleneck with increasing
intensity of the turbulence, measured by the Taylor micro-scale Reynolds
number $R_{\lambda}$. Here we report first experimental results on
the dependence of the amplitude of the bottleneck as a function of
$R_{\lambda}$ in a wind-tunnel flow. We used an active grid \cite{griffin_control_2018}
in the Variable Density Turbulence Tunnel (VDTT) \cite{bodenschatz_variable_2014}
to reach $R_{\lambda}>5000$, which is unmatched in laboratory flows
of decaying turbulence. The VDTT with the active grid permitted us
to measure energy spectra from flows of different $R_{\lambda}$,
with the small-scale features appearing always at the same frequencies.
We relate those spectra recorded to a common reference spectrum, largely
eliminating systematic errors which plague hotwire measurements at
high frequencies. The data are consistent with a power law for the
decrease of the bottleneck strength for the finite range of $R_{\lambda}$
in the experiment. 
\end{abstract}

\maketitle

\section{Introduction}

Turbulence is omnipresent in natural and technological flows. Its
consequences for the associated processes are essential in the fields
of astrophysics, geophysics, meteorology, biology, and in many engineering
disciplines from chemical engineering, combustion science, heat and
mass transfer engineering to aeronautics, marine \foreignlanguage{american}{science}
and renewable energy research. From the fundamental perspective the
mathematical field theory of the incompressible Navier Stokes equation
continues to challenges pure and applied mathematicians \cite{fefferman_existence_2006}.
In turbulence fluid velocities and accelerations fluctuate greatly
and any description can only be statistical in nature. It is believed
that at very high turbulence levels at spatial scales smaller than
the energy injection scale the turbulence shows universal properties,
independent of the particular driving. According to Kolmogorov's phenomenology
from 1941 \cite{kolmogorov_local_1941} (abbreviated K41), the universal
statistical spatial properties of fully developed turbulence can be
captured in three ranges of spatial scales. Kinetic Energy is injected
into the turbulent fluctuations at the largest scales, whose properties
are particular to the driving mechanism. The kinetic energy is transformed
into heat at the very smallest scales through viscous dissipation.
If the range of spatial scales found in the turbulent structures is
large enough, a third range of scales develops, where neither the
pecularities of energy injection, nor viscous dissipation influence
the spatial scale-to scale energy transfer. This range is called the
inertial range. In this intermediate range statistical properties
can be interpreted by the scale-to-scale transfer of kinetic energy
only, described by the kinetic energy dissipation range $\varepsilon$(dissipated
power per unit mass). The dimensionless quantity used to give the
strength of turbulence and thus the size of the inertial range scaling
is the Taylor microscale Reynolds number 
\[
R_{\lambda}=\frac{u\lambda}{\nu}.
\]
$u$ is the rms of the velocity fluctuations, $\nu$ is the kinematic
viscosity of the fluid, and $\lambda$ is the Taylor microscale, which
can be thought of as the smallest length scale at which molecular
viscosity can be neglected \cite{taylor_statistical_1935}. It can
therefore be interpreted as a typical size of an inertial range eddy.
In statistically isotropic and homogeneous turbulence $R_{\lambda}$
can be linked to the well-known Reynolds number $\text{Re}=uL/\nu$
based on the large scales $L$ via $R_{\lambda}=\sqrt{15\text{Re}}$
\cite{taylor_spectrum_1938}. The integral scale $L$ can be estimated
as the integral over the velocity correlation function $L_{11}=\int\langle u(x+r)u(r)\rangle dr$.

In K41 phenomenology for spatially homogeneous and statistically isotropic
turbulence the spatial energy spectrum in the inertial range is given
by

\begin{equation}
E(k)=C_{K}\varepsilon^{2/3}k^{-5/3}.\label{eq:K41}
\end{equation}
$C_{K}$ is the Kolmogorov constant, $k$ is the wavenumber. In this
K41 spectrum the only free parameter is the dissipation rate $\varepsilon$
as indicated above.

Despite its simplicity, Eq. (\ref{eq:K41}) describes the energy spectrum
of observed and simulated turbulent flows quite well (see \cite{saddoughi_local_1994}
for a compilation). Nevertheless, important deviations are well known.
When analyzing the compensated spectrum, $E(k)\varepsilon^{-2/3}k^{5/3}$,
deviations from a $k^{-5/3}$ scaling are found. Prominent is an increase
in amplitude of the compensated spectrum at the high-wavenumber end
of the inertial range. This pileup of energy is commonly called the
bottleneck effect \cite{verma_local_2005,frisch_hyperviscosity_2008,kurien_cascade_2004,falkovich_bottleneck_1994,verma_energy_2007,yakhot_hidden_1993}.
It has been observed in laboratory flows (e.g. \cite{she_universal_1993,saddoughi_local_1994,kang_decaying_2003,bodenschatz_variable_2014})
and direct numerical simulations (DNS) \cite{khurshid_energy_2018,ishihara_energy_2016,ishihara_energy_2005,donzis_bottleneck_2010}
alike and is typically preceded by a distinct local minimum of the
compensated spectrum. The bottleneck peak is very shallow or almost
absent in hot-wire measurements of atmospheric boundary layer turbulence
at very high $R_{\lambda}>10^{4}$\cite{gulitski_velocity_2007,sreenivasan_is_1998,tsuji_intermittency_2004}.
It is generally less pronounced in one-dimensional spectra than in
three-dimensional ones \cite{dobler_bottleneck_2003}. The effect
is also present in structure functions and influences the rapidity
of the transition between the viscous and inertial ranges in the second-order
structure function \cite{lohse_bottleneck_1995,donzis_bottleneck_2010},
hints of which can also be found in structure functions of higher
orders \cite{sinhuber_dissipative_2017}. The most extensive analysis
of the bottleneck effect has been performed by Donzis \& Sreenivasan
\cite{donzis_bottleneck_2010} on DNS at $R_{\lambda}$ up to 1000.
They found that the bottleneck effect can be characterized as the
difference between the bottleneck peak height and the level of the
preceding minimum in the compensated spectrum. They conclude that
the bottleneck effect weakens as a function of $R_{\lambda}$ and
report a scaling of $R_{\lambda}^{-0.04}$. Furthermore, they find
that the peak of the bump occurs around $k\eta\approx0.13$ in three-dimensional
spectra, independent of $R_{\lambda}$. Here $\eta$ is the Kolmogorov
length scale, where dissipative effects are expected to dominate.

From a theoretical perspective, various explanations exist for the
bottleneck effect. Falkovich \cite{falkovich_bottleneck_1994} showed
that a small perturbation to a K41 spectrum in the energy transfer
equation leads to a correction of the form $\delta E(k)=E(k)(k/k_{p})^{-4/3}\ln^{-1}(k_{p}/k)$,
where $k_{p}$ is the bottleneck wavenumber. Kurien et al. \cite{kurien_cascade_2004}
argued that the time scale of helicity can be comparable to the energy
time scale in the inertial range, where the relative helicity is already
weak. They propose that the bottleneck effect is a change in the scaling
exponent of the energy spectrum from $-5/3$ to $-4/3$. Their DNS
supports this claim as they find a corresponding scaling range in
the three-dimensional spectrum. The scaling is absent in the one-dimensional
versions of their spectra. Frisch et al. \cite{frisch_hyperviscosity_2008}
studied hyperviscous Navier-Stokes equations (Laplacian of order $\alpha\geq2$)
and attribute the bottleneck effect to an incomplete thermalization
of high-wavenumber modes in the spatial spectrum. None of these studies
directly incorporates a $R_{\lambda}$-dependence of the bottleneck
height. Verma \& Donzis \cite{verma_energy_2007} study the nonlocal
and nonlinear mode-to-mode energy transfer ina shell model of turbulence
and find that a significant portion of the energy flux away from a
wavenumber shell goes to distant shells. Thus an efficient energy
cascade requires a large inertial range. If the inertial range is
insufficient, the energy piles up at the dissipative drop-off. As
the length of the inertial range is tightly linked to $R_{\lambda}$,
this implies a dependence of the bottleneck intensity on the Reynolds
number.

In summary, the bottleneck effect has been studied systematically
in DNS and various models. Numerical simulations indicate that the
effect gets weaker with increasing $R_{\lambda}$, which is also predicted
by Verma \& Donzis \cite{verma_energy_2007} and in agreement with
atmospheric measurements at ultra-high $R_{\lambda}$, where it is
absent.

Here we present a detailed analysis of the $R_{\lambda}$-scaling
of the bottleneck effect over an unprecedented range of $R_{\lambda}$
in a well controlled laboratory flow. The analysis of the bottleneck
effect from experimental data can be demanding as systematic errors
can cloud the results. From the perspective of the measuring instrument
a small bump in the compensated spectrum is a subtle effect that occurs
at rather high frequencies not yet resolvable in PIV or PTV measurements
and very difficult to achieve in LDV. We use classical constant temperature
hot-wire anemometry (CTA) assuming Taylor frozen flow hypothesis \cite{taylor_statistical_1935}
in the Max Planck Variable Density Turbulence Tunnel (VDTT) \cite{bodenschatz_variable_2014}.
Even with very well-established hot-wire technology, subtle changes
in the energy spectrum at high frequencies can be heavily influenced
by amplification or attenuation at such frequencies (see Sec. \ref{subsec:Thermal-Anemometry}
for a review).

In this manuscript we work around those effects and investigate the
bottleneck effect from the lowest Reynolds number at which it can
be identified ($\sim200$) up to the highest $R_{\lambda}$ ever measured
in a wind tunnel flow.

The paper is organized as follows: First, we present a concise compilation
of experimental efforts to reach high $R_{\lambda}$ and describe
the Variable Density Turbulence Tunnel. We continue with a brief review
of challenges posed by constant temperature hot-wire anemometry, especially
its frequency responses. In Sec. \ref{sec:Relative-Spectra} we introduce
the relative spectra that allow us to eliminate instrumentation errors
to a large extent. Finally we report the results of our analysis and
discuss their relevance for the scaling of the bottleneck effect with
$R_{\lambda}$.

\section{Experimental Methods\label{sec:Experimental-Methods}}

\subsection{High $R_{\lambda}$ and the Variable Density Turbulence Tunnel}

Kolmogorov's 1941 predictions of universal scaling in turbulent flows
refer to the limit of large $R_{\lambda}$, such that the regimes
of energy injection and viscous dissipation are well separated \cite{kolmogorov_local_1941}.
This condition is cumbersome to achieve practically. A large separation
of scales and therefore a large $R_{\lambda}$ is found in atmospheric
flows \cite{sreenivasan_is_1998,tsuji_intermittency_2004,gulitski_velocity_2007},
where control is impossible and stationary conditions are difficult
to achieve. They are difficult to achieve in controlled laboratory
flows, where all scales can be reliably measured. To reach high $R_{\lambda}$
one can turn two knobs: the size of the energy injection scale $L$
and the dissipation scale $\eta=(\nu^{3}/\varepsilon)^{1/4}$. In
direct numerical simulations (DNS), a compromise between the size
of the periodoc box, (limiting $L$), the spatial and temporal resolution,
the convergence time, and the available resources needs to be found.
The largest $R_{\lambda}=2340$ achieved in a DNS under these constraints
to date has been performed by Ishihara \cite{ishihara_energy_2016}.
The limits of computational capabilities in terms of resolution have
been recently pointed out by Yeung et al. \cite{yeung_effects_2018}.

In a laboratory experiment the energy injection scale $L$ is limited
by the dimensions of the apparatus. Large apparatuses can be built,
e.g. the Modane wind tunnel\cite{bourgoin_investigation_2018}, but
are prohibitively expensive to operate, especially considering the
many realizations needed for dedicated statistical studies of turbulence.
To expand the inertial range the dissipative scales of size $\sim\eta$
can be decreased by lowering the kinematic viscosity $\nu$ of the
working fluid demanding a higher resolution of the measurement instrument.
Examples for experiments in liquid helium, which has an ultra-low
kinematic viscosity, are found for example in Refs. \cite{pietropinto_superconducting_2003,salort_energy_2012,rousset_superfluid_2014,saint-michel_probing_2014}.
The authors use liquid helium as working fluid in various flow configurations
and have been reported to reach $R_{\lambda}$ up to 10000. The dissipative
scales of these flows are so small that they cannot be resolved by
current technology

Our approach to create a large inertial range is to use a closed-loop
wind tunnel filled with sulfur-hexaflouride (SF$_{6}$) at pressures
up to 15 bar \cite{bodenschatz_variable_2014} - the Variable Density
Turbulence Tunnel (VDTT). With classical grids it has been shown to
create $R_{\lambda}$ up to 1600 and Kolmogorov scales $\sim10$ \textmu m,
making even the smallest spatial scales experimentally accessible
\cite{sinhuber_decay_2015}. With a specially designed automoous
active grid (see below) it is possible to increase the energy injection
scale and thus the inertial range. As $R_{\lambda}\sim(L/\eta)^{3/4}$,
the VDTT features two independent handles to change $R_{\lambda}$
- pressure and active grid forcing. In combination they create a laboratory
flow of $R_{\lambda}$ more than 5000 at scales resolvable with modern
thermal anemometry under the limitations described below.

\begin{figure}
\includegraphics[width=1\columnwidth]{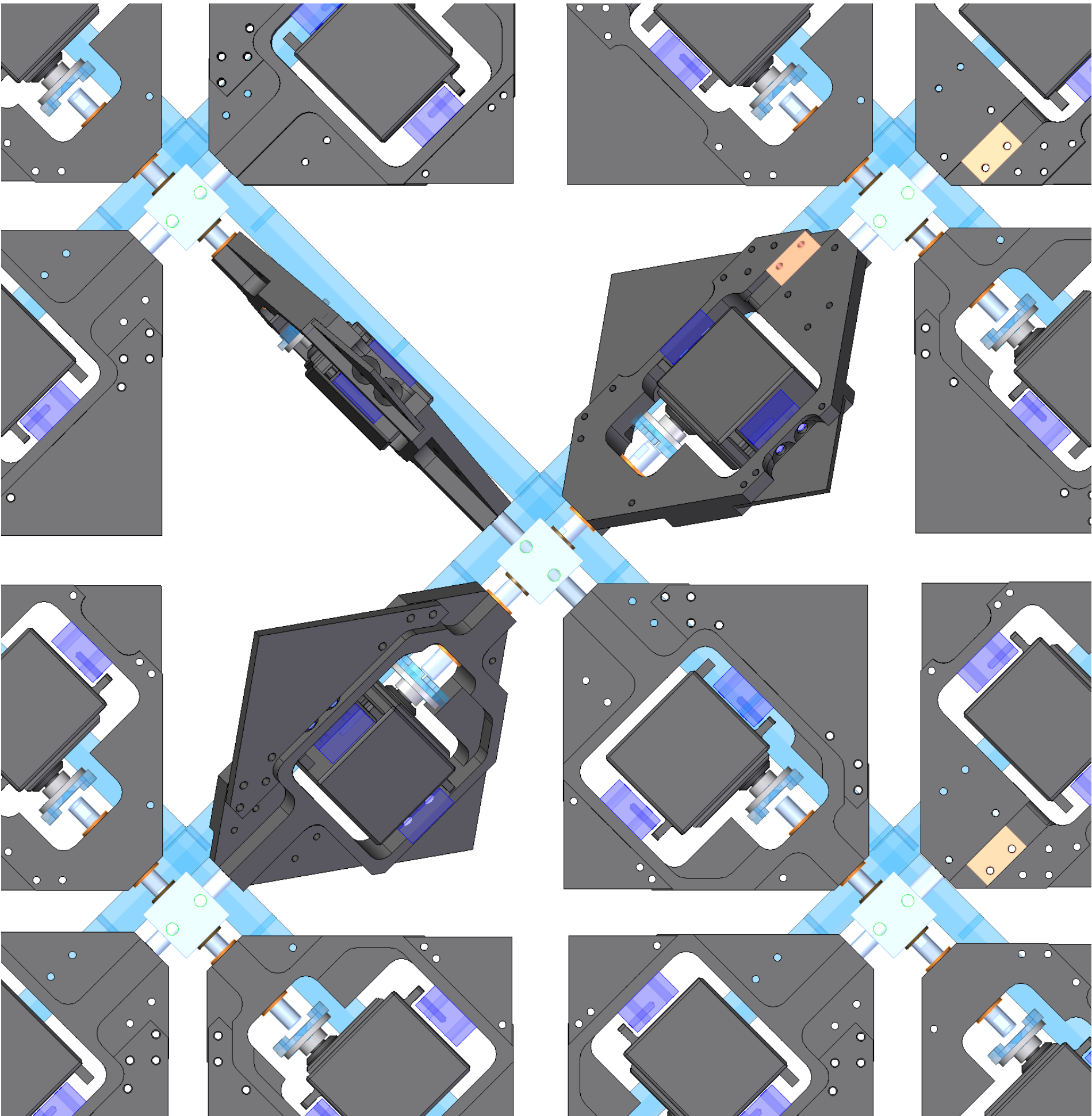}

\caption{Several flaps of the active grid. The flow points out of the page.
Starting from the top left flap in clockwise direction the flaps are
set to 0º, 45º, 90º, and 45º. The side length of one flap is 11 cm,
the black boxes in the flap center are servo motors, the blue rods
are the grid support. }
\end{figure}

The autonomous active grid consists of 111 individually controllable
flaps of dimensions 11 cm x 11 cm that rotate around their diagonal.
This is different from the Makita-style grids, where the rows and
columns of the flaps are mounted rigidly on rotating horizontal and
vertical bars \cite{hideharu_realization_1991}. The angle of rotation
can be set to any angle between $\pm90{^\circ}$. The flow obstruction
is smallest (flap parallel to the flow) at $0\text{°}$. At angles
$\neq0\text{°}$ one of the flap sides is facing the incoming flow,
while the other side is facing away from the flow. The sign of the
angles determines the side that is facing the flow, while the magnitude
defines the deviation of the flap from the parallel position. As in
a classical grid with rigid grid bars, wakes are formed that interact
with each other downstream of the grid to form a turbulent flow field.
The flexibility of the grid allows the superposition of larger structures
onto those induced by the individual flaps. A detailed account of
the autunomous active grid and the algorithm is given in Ref. \cite{griffin_control_2018}
and briefly summarized here.

The algorithm updates the angle of each flap every 0.1s. Each time
step starts with a random set of angles and convolves each of those
angles with the history and a pre-defined kernel. The kernel is always
defined by a certain shape (e.g. Gaussian), the spatial and temporal
correlations (the number of neighbors and time-steps included in the
convolution), and the desired mean absolute angle $\phi_{RMS}$. More
complex kernel shapes require additional parameters. For the experiments
presented here, a 'Long Tail' kernel has been used, which reduces
the correlation with neighboring flaps and therefore emphasizes the
correlation of the angle with its own past.

\begin{figure}
\includegraphics[width=1\columnwidth]{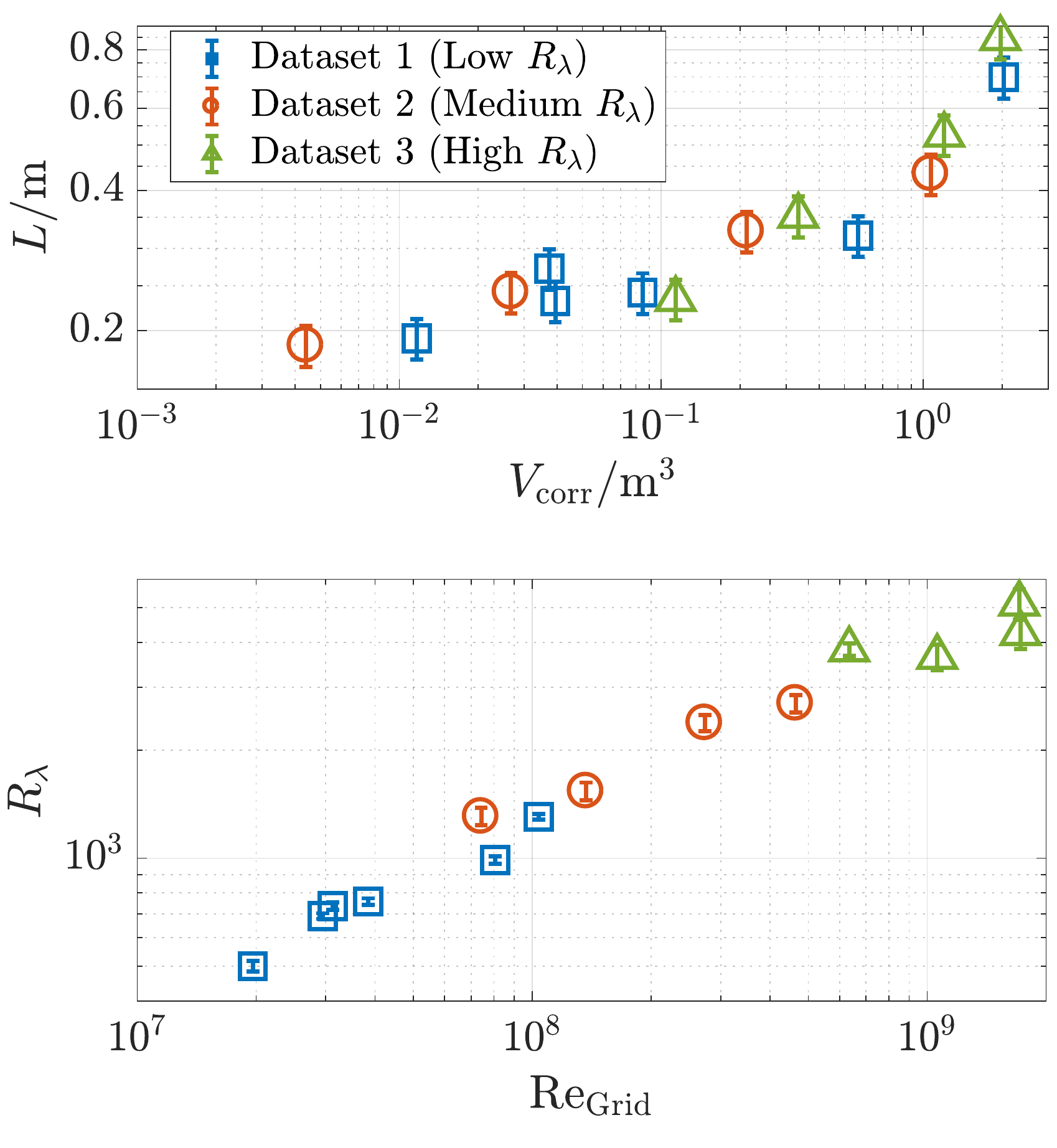}

\caption{(a) The correlation volume $V_{\text{corr}}=\sigma_{s}^{2}\sigma_{t}U$
influences the size of the largest scales $L$. (b) The grid Reynolds
number defined as $\text{Re}_{\text{Grid}}=\sqrt[3]{V_{\text{corr}}}\phi_{RMS}U/\nu$
determines $R_{\lambda}$. The flattening of the individual Datasets
indicate a limit to the increase of $R_{\lambda}$ through the active
grid. The first spectrum of Dataset 1 is not shown, because $V_{\text{Corr}}$
is are not defined for a stationary, open grid. \label{fig:Mean-grid-flap}}
\end{figure}

This algorithm leads to dynamically evolving patches of more open
and more closed flaps without periodicity. The typical time- and length
scales of those patches are controlled by the spatial and temporal
correlation lengths, $\sigma_{s}$ and $\sigma_{t}$, respectively,
and the mean flap angle $\phi_{RMS}$ defines their mean amplitude.
We describe the correlation as a box of dimensions $\sigma_{s}\times\sigma_{s}\times(\sigma_{t}\cdot U)$
and relate this volume $V_{\text{Corr}}$ to the energy injection
scales $L=\int C(r)dr$ in Fig. \ref{fig:Mean-grid-flap} a. As expected,
larger correlation volumes lead to larger energy injection scales,
with a corresponding change in $R_{\lambda}$. $\sigma_{s}$ and $\sigma_{t}$
can be set independently, but to avoid a strongly inhomogeneous flow
they are typically linked via the mean flow velocity $U$ forming
a cubic correlation box: $\sigma_{t}\cdot U\approx\sigma_{s}$. This
rule-of-thumb needs to be relaxed slightly to achieve $R_{\lambda}>3000$
leading to a correlation box that is elongated in the $\sigma_{t}$
direction.

The grid parameters can be distilled further into a grid Reynolds
number. The relevant length scale is given by $\sqrt[3]{V_{\text{Corr}}}$,
which corresponds to $\sigma_{s}$ in the case of a cubic correlation
box. The fluctuating velocity is proportional to the mean flow velocity
and the mean angle amplitudes $\phi_{RMS}$: 
\[
\text{Re}_{\text{Grid}}\sim\frac{\sqrt[3]{V_{\text{Corr}}}(\phi_{RMS}U)}{\nu},
\]
where $U$ is the mean flow velocity of the VDTT and $\nu$ the kinematic
viscosity. Fig. \ref{fig:Mean-grid-flap} b) shows that the \textit{a
priori} quantity $\text{Re}_{\text{Grid}}$ scales with the \textit{a
posteriori $R_{\lambda}$ }with deviations at $\text{Re}_{\text{Grid}}>10^{9}$.
Each Dataset has been obtained by increasing $V_{\text{Corr}}$ while
keeping the pressure (and therefore $\nu$ constant) as indicated
in Tab. \ref{tab:Summary-of-Datasets.}. We attribute the slight deviations
at large $\text{Re}_{\text{Grid}}$ from a power law dependence to
the fact that $L$ is approaching half the diameter of the measurement
section. This is a natural limit for a sensible energy injection in
any tunnel. We would like to add the word of caution that when approaching
this limit, isotropy and homogeneity cannot be assumed easily anymore,
which leads to said deviations from the isotropic relation $R_{\lambda}\sim\text{Re}_{\text{Grid}}^{\zeta}$
with $\zeta\approx0.5$. Nevertheless, these data confirm that the
active grid is indeed another 'knob' to change $R_{\lambda}$ through
the large scales.

\subsection{Thermal Anemometry\label{subsec:Thermal-Anemometry}}

More than a century after its invention \cite{comte-bellot_hot-wire_1976},
hot-wire anemometry remains the technique of choice to measure the
energy spectrum of turbulent velocity fluctuations in a strong mean
flow. Constant temperature anemometry is responsive to fluctuations
up to very high frequencies. The sensing element's resistance - and
therefore its temperature - is kept constant by a feedback circuit.
As long as the feedback circuit is fast enough, the thermal lag of
the wire does not attenuate fluctuations faster than the thermal time
scale of the wire. This comes at the expense of a mores complicated
circuitry and frequency response.

The frequency response of CTA circuits has been studied extensively
both through theoretical models and experimental testing. Freymuth\cite{freymuth_frequency_1977}
linearized a circuit with a single feedback amplifier of infinitely
flat frequency response and analyzed its response to square and sine
waves. He finds that the system can be modeled by a third-order ODE
if the circuit responds faster than the wire, and the frequency response
is optimal (flat over the entire range of frequencies) when the system
response to a step perturbation by a single, slight overshoot (critically
damped system). Perry \& Morrison \cite{perry_study_1971} investigated
more moderate amplifier gains and bridge imbalances in their study
yielding similar results. Wood \cite{wood_method_1975} expanded
the Perry \& Morrison analysis, but considered a single-stage amplifier
with a frequency-dependent response. Watmuff \cite{watmuff_investigation_1995}
further expanded the model with multiple, non-ideal amplifier stages.
He showed that at least two amplifier stages are necessary to model
the real amplifier properly. This introduces two additional poles
to the system and makes the frequency response more complicated. Samie
et al. \cite{samie_modelling_2016} recently studied anemometry with
sub-miniature probes in this model and compared it to a real CTA measurement.
The results supported the further development of their in-house circuit,
such that sub-miniature hot wire probes could be operated successfully
on this CTA for the first time.

These theoretical attempts to predict the frequency response of a
CTA circuit are accompanied by experimental approaches. Bonnet and
de Roquefort \cite{bonnet_determination_1980} heated the wire periodically
by a perturbation voltage as well as laser heating to determing the
frequency response. Weiss et al. \cite{weiss_method_2001} used the
aforementioned square wave test and interpreted its power spectrum
as a measure for the frequency response curve. Hutchins et al. \cite{hutchins_direct_2015}
exploited the well-defined frequency content of pipe flow at different
operating pressures to obtain frequency response curves without artificial
heating. They were able to create flows of almost identical Reynolds
number, but different frequency content and could deduce the frequency-response
curves for different circuits and wires. They compared several anemometer
circuits and wires and found that the frequency responses are non-constant
at frequencies as low as 500 Hz. For the combination of CTA circuit
and wire used in the present study, they report an attenuation between
400 Hz and 7 kHz followed by a strong amplification of the signal.
We therefore cannot assume a flat frequency response for our measurements
and adress these effects below.

The energy spectrum measured by a hot wire is influenced by the effects
of finite wire length. Length scales smaller than the sensor's sensing
lengths $l$ will be attenuated, but also larger wavenumbers are influenced.
Wyngaard \cite{wyngaard_measurement_1968} used a Pao model spectrum
\cite{pao_structure_1965} to investigate this attenuation of small
scales. These results were reviewed in Ref. \cite{mckeon_velocity_2007}
indicating that for $l/\eta=2$, the attenuation of the one-dimensional
spectrum is still minimal at $k\eta\sim0.3$, which was supported
Ashok et al. \cite{ashok_hot-wire_2012}. Sadeghi et al. \cite{sadeghi_effects_2018}
used sub-miniature hot wires (NSTAPs) as a benchmark and found that
spatial filtering of the energy spectrum is minimal for $l/\eta<3.7$
at $k\eta<0.1$.

In this study we used conventional hot wires of sensing length 450
\textmu m for pressures below 2 bar, as well as Nanoscale Thermal
Anemometry Probes (NSTAP) of sensing length 30 \textmu m provided
by Princeton University with a Dantec Dynamics StreamWare CTA circuit.
The NSTAP is a 100 nm thick, 2.5 \textmu m wide, and 30 or 60 \textmu m
long free-standing platinum film supported by a silicon structure
and soldered to the prongs of a Dantec hot wire. The production process
and characteristics are detailed in Refs. \cite{fan_nanoscale_2015,kunkel_development_2006,bailey_turbulence_2010,vallikivi_fabrication_2014}.
For the conventional hot wire $l/\eta<5$ in all cases and for the
NSTAP $l/\eta<3$. Therefore, $\eta$ cannot be fully resolved in
all cases. However, the bottleneck effect is typically found around
$100\eta$. The aforementioned references show that we can regard
the distortions due to finite wire length as minor in this part of
the energy spectrum.

To summarize, the spatial resolution of our measurement instruments
is sufficient to study the $R_{\lambda}$-dependence of the bottleneck
effect. Nevertheless the nonlinear frequency response of the circuitry
remains. Here we describe a procedure that takes the response into
account and thus removes this systematic measurement error.

\section{Relative Spectra\label{sec:Relative-Spectra}}

\begin{table*}
\begin{centering}
\begin{tabular}{|c|c|c|c|c|c|c|cc|}
\hline 
Dataset  & $R_{\lambda}$  & $p$(bar)  & $U$(m/s)  & $u$(m/s)  & Sensor  & $L=\int C(r)dr$ (m)  & $\eta$ (\textmu m)  & $f_{\eta}$(kHz)\tabularnewline
\hline 
\hline 
1  & 193  & 1.5  & 2.75  & 0.04  & Regular HW  & 0.12  & 277  & 9.9\tabularnewline
\hline 
1  & 500  & 1.5  & 2.41  & 0.14  & Regular HW  & 0.19  & 139  & 17.3\tabularnewline
\hline 
1  & 690  & 1.5  & 2.41  & 0.18  & Regular HW  & 0.23  & 127  & 19.0\tabularnewline
\hline 
1  & 735  & 1.5  & 2.65  & 0.19  & Regular HW  & 0.27  & 127  & 20.9\tabularnewline
\hline 
1  & 757  & 1.5  & 2.43  & 0.20  & Regular HW  & 0.24  & 116  & 20.9\tabularnewline
\hline 
1  & 989  & 1.5  & 2.72  & 0.22  & Regular HW  & 0.32  & 120  & 22.7\tabularnewline
\hline 
\textbf{1}  & \textbf{1305}  & \textbf{1.5}  & \textbf{2.28}  & \textbf{0.34}  & \textbf{Regular HW }  & \textbf{0.70}  & \textbf{91}  & \textbf{25.1}\tabularnewline
\hline 
\hline 
\rowcolor[gray]{0.95}\textbf{2}  & \textbf{1308}  & \textbf{5.95}  & \textbf{3.64}  & \textbf{0.20}  & \textbf{NSTAP 1 }  & \textbf{0.19}  & \textbf{37}  & \textbf{98.4}\tabularnewline
\hline 
\rowcolor[gray]{0.95}2  & 1539  & 5.95  & 3.68  & 0.22  & NSTAP 1  & 0.24  & 36  & 102.2\tabularnewline
\hline 
\rowcolor[gray]{0.95}2  & 2385  & 5.97  & 3.64  & 0.35  & NSTAP 1  & 0.33  & 28  & 130.0\tabularnewline
\hline 
\rowcolor[gray]{0.95}2  & 2704  & 5.97  & 3.61  & 0.40  & NSTAP 1  & 0.43  & 26  & 138.9\tabularnewline
\hline 
\hline 
3  & 3641  & 14.62  & 3.75  & 0.44  & NSTAP 2  & 0.38  & 10  & 375.0\tabularnewline
\hline 
3  & 3821  & 14.71  & 3.83  & 0.41  & NSTAP 2  & 0.27  & 12  & 319.2\tabularnewline
\hline 
\textbf{3}  & \textbf{4247}  & \textbf{14.65}  & \textbf{3.97}  & \textbf{0.53}  & \textbf{NSTAP 2}  & \textbf{0.54}  & \textbf{9.2}  & \textbf{431.5}\tabularnewline
\hline 
3  & 5130  & 14.66  & 4.01  & 0.58  & NSTAP 2  & 0.86  & 9.1  & 440.7\tabularnewline
\hline 
\end{tabular}
\par\end{centering}
\caption{Properties of all spectra. All spectra of a Dataset are distorted
by the same function $T(f)$ describing the sensor- and instrument-induced
bias. This is ensured by changing $R_{\lambda}$ only through large
scales $L$ and fixing the position of the small scales in frequency
space indicated by $f_{\eta}$. A reference spectrum has been chosen
from each dataset, which is emboldened in this table. \label{tab:Summary-of-Datasets.}}
\end{table*}

\subsection{The concept}

As outlined above, systematic errors influence the energy spectra
recorded with a hot-wire anemometer as outlined above. Formally, this
means that the one-dimensional energy spectrum $E_{11}(f)$ is distorted
by a frequency-dependent transfer function $T(f)$: 
\[
E_{M}(f)=E_{11}(f)\hphantom{}T(f)
\]
$T(f)$ describes the effects of the thermal wire response, which
depends on pressure and speed, and the reponse of the constant temperature
anemometry circuit. Ideally, $T(f)$ is a constant over the whole
range of relevant frequencies, but the evidence detailed above indicates
a complex shape of amplification and attenuation of the signal. In
this study we do not make any attempt to find $T(f)$. Instead, we
control its effects by keeping $T(f)$ the same for several flows
at different $R_{\lambda}$.

To ensure that the spectra only differ because of changes in the turbulent
fluctuation and not because of the frequency response curve of the
anemometer, we need to ensure that the response curve $T(f)$ is unaltered
between spectra. We achieve this in two steps. The ambient pressure
might influence the heat transfer of the wire and therefore $T(f)$.
Furthermore, $T(f)$ is influenced by the CTA tuning (in particular
the overheat), and the sensor itself. Therefore, we fix the ambient
pressure within a set of spectra (a 'Dataset') and measure using the
same sensor and the same CTA settings.

The second step is to ensure that a given $k\eta$ is influenced by
the same part of the frequency response curve $T(f)$. Thus, we need
to fix the position of a spectral feature in frequency space. This
means that the mean velocity $U$ must be the same within one Dataset.
$T(f)$ mainly distorts the small-scale end of the spectrum \cite{freymuth_frequency_1977,hutchins_direct_2015,mckeon_velocity_2007,perry_study_1971,samie_modelling_2016,watmuff_investigation_1995,weiss_method_2001,wood_method_1975},
whose location in frequency space at a given $U$ is determined by
the kinematic viscosity $\nu$. $\nu$ is fixed within a Dataset because
the pressure remains constant.

We can, however, change the energy injection scale and thus the $R_{\lambda}$
with the autonomous active grid. This way we can conduct measurements
at different $R_{\lambda}$. Ultimately, we can eliminate $T(f)$
by relating each spectrum to a reference spectrum: 
\begin{align}
\frac{E_{M}^{i}(f)}{E_{M}^{\text{Ref}}(f)} & =\frac{E_{11}^{i}(f)T^{i}(f)}{E_{11}^{\text{Ref}}(f)T^{\text{Ref}}(f)}\nonumber \\
 & =\frac{E_{11}^{i}(f)T(f)}{E_{11}^{\text{Ref}}(f)T(f)}=\frac{E_{11}(kU/2\pi)}{E_{11}^{\text{Ref}}(kU/2\pi)}.
\end{align}
In the following we call the ratio of a spectra divided by a reference
spectrum in the frequncy domain, relative spectrum.

\subsection{Results}

We created three sets of spectra that have identical $T(f)$ each.
We call these sets `Datasets`. Tab. \ref{tab:Summary-of-Datasets.}
shows important parameters for each spectrum. Note that $L$ changes
significantly within a given dataset leading to changes in $R_{\lambda}$,
while $f_{\eta}=U/\eta$ remains almost constant within the dataset.
This indicates that we changed $R_{\lambda}$ only by increasing the
large scales, while keeping all small-scale features of the spectrum
at the same frequency $f_{F}$. For example, in Dataset 2, the peak
of the spectral bump always lies at a frequency of $\sim700$ Hz,
whereas the beginning of the inertial range spans a factor of 4 in
frequency (2 to 8 Hz). This exemplifies the excellent control over
$R_{\lambda}$ permitted by the autonomous active grid as indicated
in Fig. \ref{fig:Mean-grid-flap}.

\begin{figure}
\includegraphics[width=1\columnwidth]{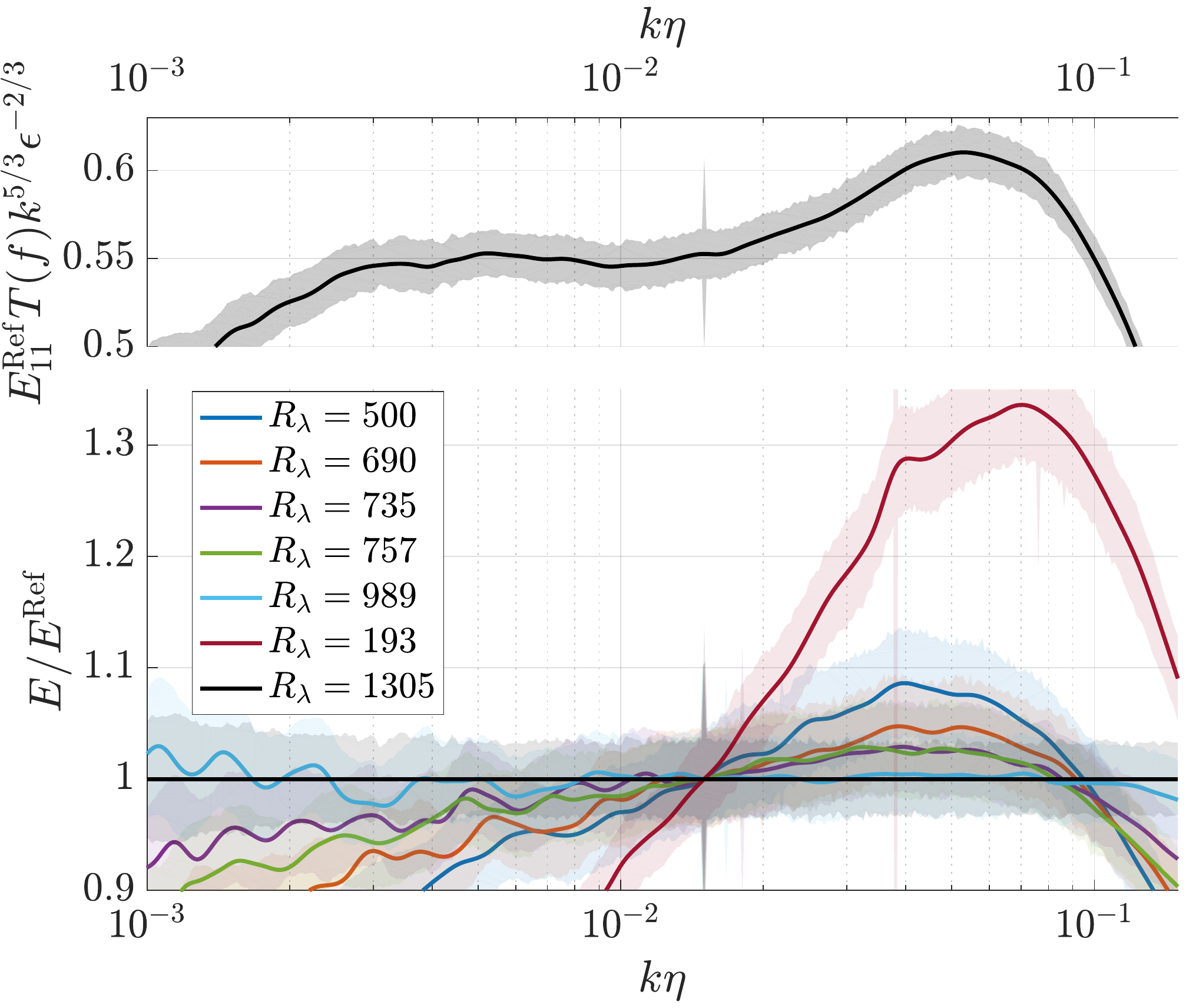}

\caption{Reference spectrum at $R_{\lambda}=989$ (upper plot) and relative
spectra from Dataset 1. The data have been collapsed at $k\eta=0.015$,
which we defined as the beginning of the bottleneck region. We identified
the peaks in the relative spectra with the bottleneck peak of the
absolute spectra. The peak height decreased with increasing $R_{\lambda}$
and different spectra of similar $R_{\lambda}$ result in very similar
relative spectra as expected. Furthermore, the slope of the spectrum
at $k\eta<0.015$ seems to decrease with $R_{\lambda}$. The shaded
areas are a measure of the noise level. \label{fig:LowReScan}}
\end{figure}

\begin{figure}
\includegraphics[width=1\columnwidth]{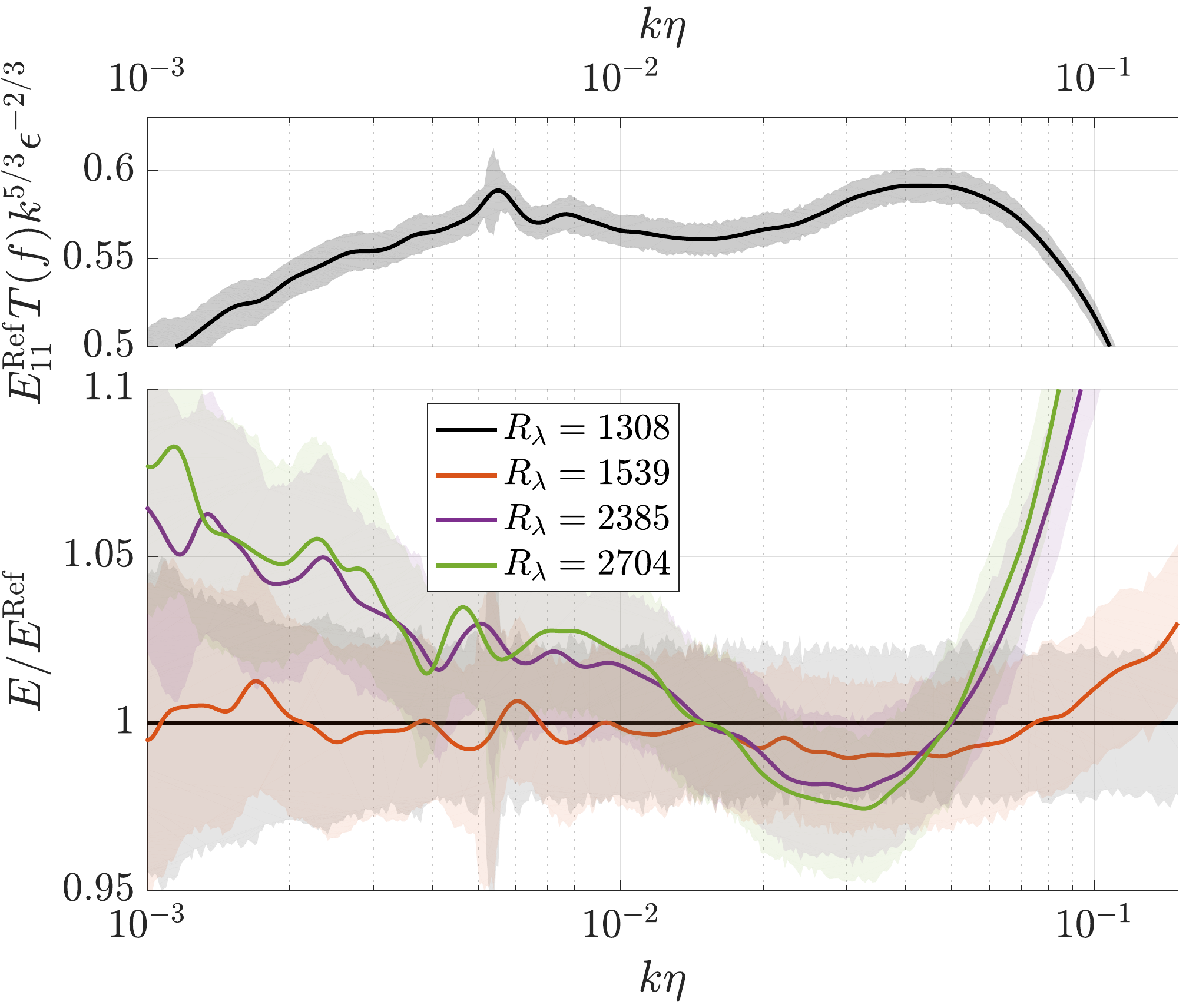}

\caption{Reference spectrum at $R_{\lambda}=1308$ (upper plot) and relative
spectra from Dataset 2. The trends in peak height and slope from Fig.
\ref{fig:LowReScan} continue. \label{fig:MediumReScan}}
\end{figure}

\begin{figure}
\includegraphics[width=1\columnwidth]{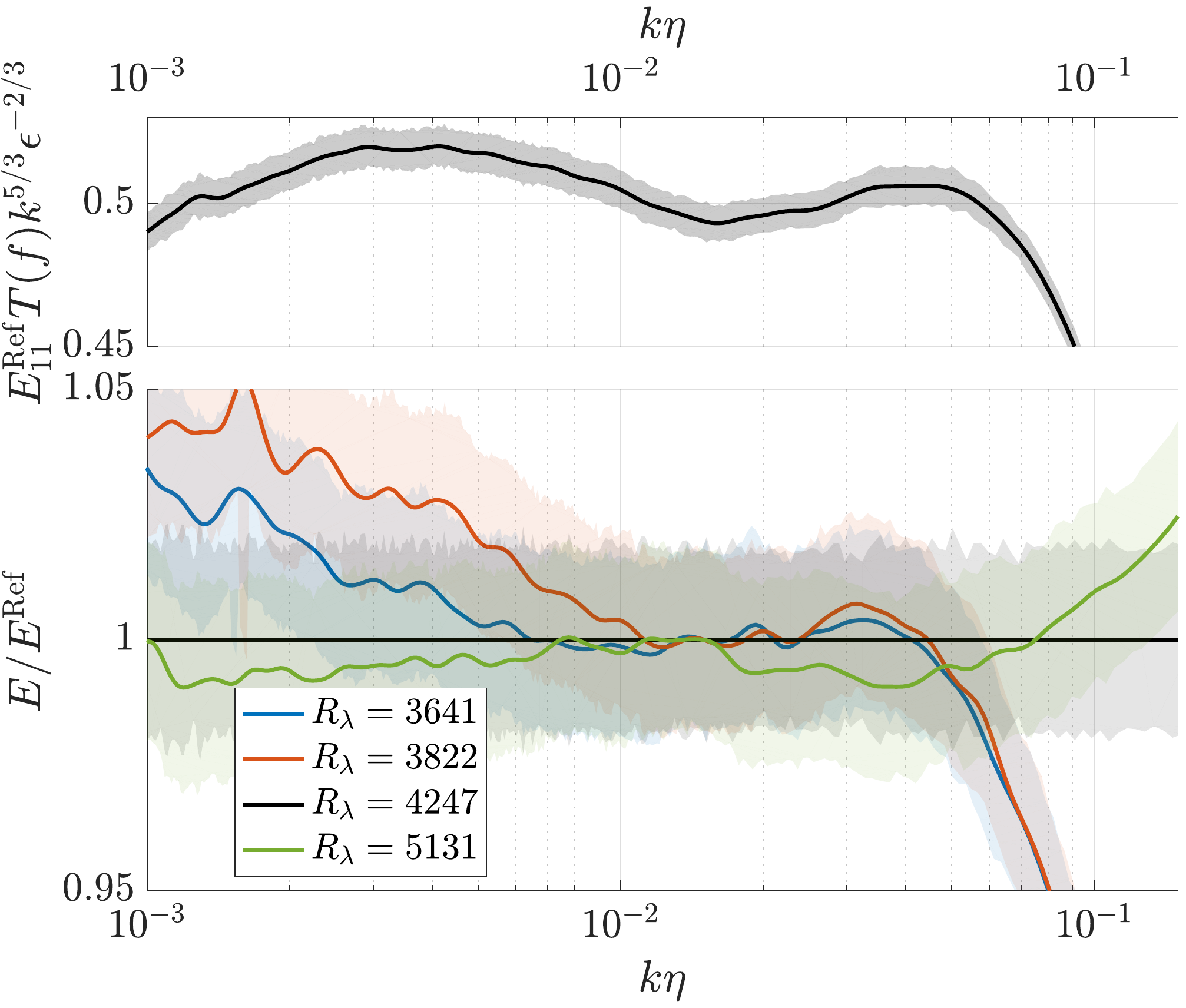}\caption{Reference spectrum at $R_{\lambda}=4247$ (upper plot) and relative
spectra from Dataset 3. Unlike in Datasets 1 and 2, the beginning
of the bottleneck region around $k\eta=0.015$ is identifiable in
the relative spectra as a local extremum. \label{fig:HighReScan}}
\end{figure}

The lower graphs of Figs. \ref{fig:LowReScan} - \ref{fig:HighReScan}
show the spectra from each of the respective datasets divided by the
reference spectrum $E_{11}^{\text{Ref}}$. $E_{M}^{\text{Ref}}$ is
plotted pre-compensated in the upper graphs of the respective figure.
Note that the absolute spectra in the upper graphs are multiplied
by an unknown transfer function $T(f)$ accounting for probe effects
and therefore can not be used to realibly measure the features of
the bottleneck. However, the relative curves are corrected and allow
a measurement. The graphs are the result of a smoothing procedure
and error estimate detailed in the appendix. In brief, the spectra
were smoothed using a $1/f$ octave filtering and the error is related
to the noise level removed by the smoothing procedure. The spectra
have been divided by the reference spectrum in the frequency domain
and collapsed at $k\eta=0.015$ afterwards to simplify interpretation.

While in Dataset 3 the beginning of the bottleneck region around $k\eta=0.015$
is accompanied by a change in the shape of the relative spectra, this
point cannot be identified in the relative spectra of Datasets 1 and
2. The relative spectra seem to follow approximately straight lines
in our semilogarithmic plot, i.e. $E_{11}/E_{11}^{\text{Ref}}\sim\zeta(R_{\lambda})\log(k\eta)$.
The slope of these lines appears to become less steep with $R_{\lambda}$,
leading to the prefactor $\zeta(R_{\lambda})$. In the following we
concentrate on the bottleneck effect found at $k\eta>0.015$ for the
remainder of this section.

The location of the spectral bump forming the bottleneck effect in
relative spectra is not obvious. However, when considering the background
noise, the peak location is not the major source of error. E.g. for
$R_{\lambda}=1539$, all points between $0.015<k\eta<0.07$ are within
the errorband at $k\eta=0.05$. We therefore define the extremum in
the relative spectrum between $0.015<k\eta<0.08$ as the relative
height $h$ of the bottleneck effect. This has the additional advantage
to be independent of the errors in the estimate of $\eta$. To preclude
biases from this definition, we repeat our analysis with different
definitions of the relative bottleneck height in Fig. (\ref{fig:DifferentBNDefs})
in the appendix.

Finally, the measured bottleneck height cannot depend on which spectrum
is chosen as reference. We have calculated the bottleneck height with
all possible choices of $E_{11}^{\text{Ref}}$ and found our results
to be largely independent of that choice (see Appendix for details).

Fig. \ref{fig:Bottleneck-height-RelativeRe} shows the bottleneck
height - defined as above - as a function of \LyXZeroWidthSpace $R_{\lambda}/R_{\lambda}^{\text{Ref}}${}
within each dataset . The data shows a trend towards smaller peak
heights in the relative spectrum with increasing $R_{\lambda}$. The
data follows the numerical data we have compiled from various sources
\cite{buaria_characteristics_2015,yeung_extreme_2015,buaria_extreme_2018,ishihara_energy_2016}.
We have analyzed the data from Buaria et al. \cite{buaria_extreme_2018}
at $R_{\lambda}$ up to 650 ($R_{\lambda}/R_{\lambda}^{\text{Ref}}<1$).
The increased small-scale resolution in comparison to \cite{donzis_bottleneck_2010}
seems to have no noticable impact on the bottleneck. Therefore, this
data at is practically the same as the one used by Donzis \& Sreenivasan
\cite{donzis_bottleneck_2010} for our purposes. The data from $R_{\lambda}=1300$
($R_{\lambda}/R_{\lambda}^{\text{Ref}}=1)$ was reported in Ref. \cite{buaria_characteristics_2015}.
The numerical data at $R_{\lambda}/R_{\lambda}^{\text{Ref}}\approx1.9$,
which corresponds to $R_{\lambda}=2340$, is the highest $R_{\lambda}$
reported by Ishihara et al. \cite{ishihara_energy_2016}. The relative
spectra of the numerical data were analyzed equivalently to the experimental
data and the spectrum at $R_{\lambda}=1300$ was chosen as a reference
spectrum.

When excluding the lowest $R_{\lambda}$, the experimental data is
in agreement with the power law of 
\[
h\sim\left(R_{\lambda}/R_{\lambda}^{\text{Ref}}\right)^{-0.061\pm0.007}.
\]
The fit was obtained by a bootstrap procedure based on the error bars.

The spectrum at $R_{\lambda}=193$ follows the general trend of decreasing
peak height with $R_{\lambda}$, but its peak differs substantially
from the predictions. The absolute spectrum (not shown) exhibits no
signs of a $5/3$-scaling, and consequently the bottleneck region
cannot be clearly separated from the rest of the spectrum. This is
substantially different from the other spectra, where the end of the
integral range could always be observed in the absolute spectra and
we therefore are not surprised that the relative spectrum at $R_{\lambda}=193$
deviates from the remainder of the data. This spectrum has consequently
been ignored in our interpretation.

Further, we can change $R_{\lambda}$ only by a factor of 5 through
the autonomous active grid. While Dataset 1 and 2 each feature a spectrum
at the same $R_{\lambda}$, there is a gap between the highest $R_{\lambda}$
of Dataset 2 (2704) and the lowest of Dataset 3 (3641). To plot $h$
as a function of $R_{\lambda}$ alone, we use the aforementioned power
law fit from Fig. \ref{fig:Bottleneck-height-RelativeRe}, i.e. we
assume $h(R_{\lambda}=3641)=h(R_{\lambda}=3641)(3641/2704)^{-0.061}$
to arrive at Fig. \ref{fig:Bottleneck-height-relative}.

\begin{figure}
\includegraphics[width=1\columnwidth]{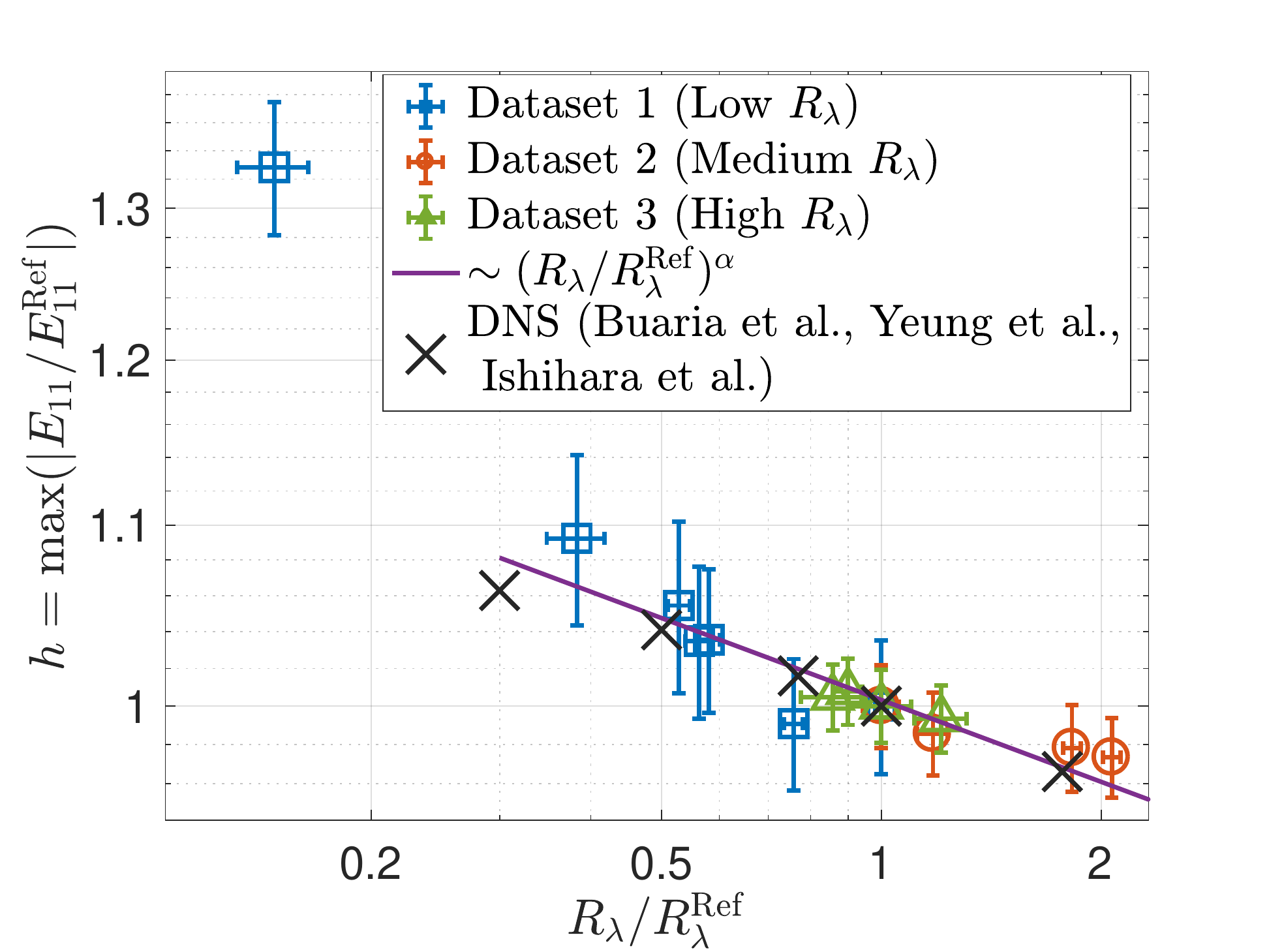}

\caption{Bottleneck height relative to the reference spectrum for all datasets
as a function of $R_{\lambda}/R_{\lambda}^{\text{Ref}}$. The reference
spectra have bottleneck height 1. Numerical simulations from Buaria
et al. \cite{buaria_extreme_2018} at $R_{\lambda}$ up to 650 ($R_{\lambda}/R_{\lambda}^{\text{Ref}}<1$),
Buaria et al. \cite{buaria_characteristics_2015} ($R_{\lambda}=1$)
and Ishihara \cite{ishihara_energy_2016} ($R_{\lambda}>1$ ) are
added for reference. A power law is fitted to the experimental data
with the lowest $R_{\lambda}/R_{\lambda}^{\text{Ref}}$ excluded.
$(R_{\lambda}/R_{\lambda}^{\text{Ref}})^{-0.061\pm0.007}$ is a good
description of the experimental data over one decade of $R_{\lambda}$
(from 500 to 5000) and agrees with the numerical simulations as well.
This power law is used in Fig. \ref{fig:Bottleneck-height-relative}
to combine Datasets 2 and 3. \label{fig:Bottleneck-height-RelativeRe}}
\end{figure}

\begin{figure}
\includegraphics[width=1\columnwidth]{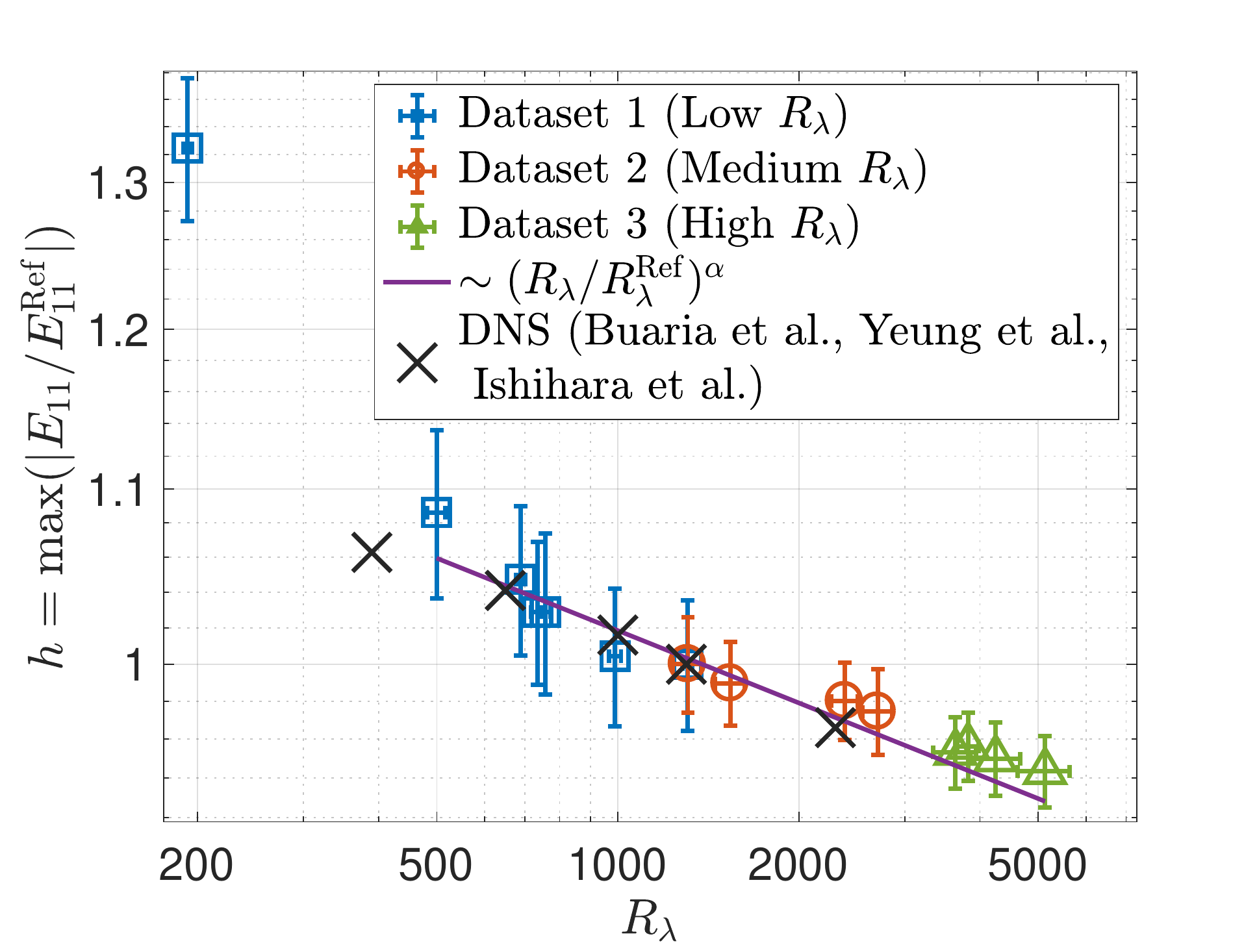}

\caption{We have shifted Dataset 3 from Fig. \ref{fig:Bottleneck-height-RelativeRe}
under the assumption of a power law $\sim\left(R_{\lambda}/R_{\lambda}^{\text{Ref}}\right)^{\alpha}$
with $\alpha=-0.061\pm0.007$, i.e. the position of Dataset 3 with
respect to the other Datasets is constructed artificially from the
power law in Fig. \ref{fig:Bottleneck-height-RelativeRe}. We have
no physical justification for this power law and stress that the position
of Dataset 3 in this figure is speculative. \label{fig:Bottleneck-height-relative}}
\end{figure}

\section{Discussion}

In this paper we studied the spectra of a turbulent wind tunnel flow
of $R_{\lambda}$ between 193 and 5131. We have used regular hot-wires
as well as NSTAPs with a state-of-the-art constant temperature anemometer
to record single-\LyXZeroWidthSpace point two-time statistics of the
turbulent fluctuations, in particular energy spectra. However, such
spectra can be heavily influenced by non-ideal frequency responses
of the circuit. The frequency response is particularly complicated
when operating sub-miniature wires like the NSTAP with a CTA \cite{samie_modelling_2016,hutchins_direct_2015}.
A constant current \LyXZeroWidthSpace anemometer (CCA) might perform
better in this respect, because the frequency reponse is limited only
by the thermal lag of the wire and no feedback loop is involved. Still,
this comes at the expense of a variable wire temperature and -resistance.

In an attempt to interpret CTA data suffering from a non-flat frequency
response, we consider energy spectra relative to a reference spectrum.
Such an analysis significantly restricts the phenomena that can be
observed. The bump in the energy spectrum at the transition from the
inertial to the dissipation range can still be identified in the relative
spectra as a local extremum beyond $k\eta=0.015.$

To the best of our knowledge, no other wind tunnel achieves $R_{\lambda}>5000$
in a gas. Moreover, we do not know of any other quantitative study
of the scaling of the bottleneck effect with $R_{\lambda}$ in a laboratory
experiment. We attribute this to the difficulties one faces when interpreting
energy spectra from CTA measurements at relatively high frequencies:
The spectrum is stronlgy influenced by the CTA circuitry and these
influences are hard to quantify or eliminate.

With the aforementioned procedures we are able to extract information
about the bottleneck effect from instrument-distorted hot wire spectra.
We find indications that the bottleneck effect decreases up to $R_{\lambda}\sim5000$.
We fit a power law of \LyXZeroWidthSpace $(R_{\lambda}/R_{\lambda}^{\text{Ref}})^{-\alpha}${}
with $\alpha=0.061\pm0.007$, which is close to the value of $(R_{\lambda})^{-0.04}$
found by Donzis \& Sreenivasan \cite{donzis_bottleneck_2010}. Their
numerical results are in general in good agreement with our experimental
data, lending support to the experiment and data anlysis procedure.
Our data equally supports Verma \& Donzis \cite{verma_energy_2007},
who predict that the bottleneck scales as $h\sim1-\gamma(1.5\log_{2}(R_{\lambda}))^{2/3}$.

We attempt to plot the relative bottleneck height as a function of
$R_{\lambda}$ alone. This requires the assumption that the aforementioned
power law holds and can be extrapolated. Such an assumption is highly
speculative and the results should be considered as such.

We can not quantify the absolute height of the bottleneck bump. Yet,
we can argue that if the relative spectra are still changing with
$R_{\lambda}$ in the relevant region, the effect has not completely
vanished. We can find a systematic decrease of the peak in the relative
spectra for $R_{\lambda}<3000$. The data for $R_{\lambda}>3000$
in Dataset 3 is inconclusive. A small, decreasing trend can be found,
consistent with the power law fit. However, the differences in height
are so small compared to the error bars that the claim of a constant
bottleneck height at $R_{\lambda}>3000$ would also be supported by
the data, especially when considering alternative definitions of the
bottleneck height in relative spectra as in Fig. \ref{fig:DifferentBNDefs}
found in the appendix. This is not in contradiction to the atmospheric
spectra mentioned above, as they have an even higher $R_{\lambda}$.
Further, we note that a bottleneck effect might not show up as a peak
in a $5/3$-compensated spectrum, yet might be present when compensating
by an intermittency-corrected slope $-(5/3+\beta)$. In this case,
the bottleneck effect would still be visible in the relative spectrum.
However, the claim that the bottleneck height does not change with
$R_{\lambda}$ for $R_{\lambda}>3000$ is not ruled out by the data.

As far as this study is concerned, the data matches the predictions
of Verma \& Donzis \cite{verma_energy_2007}: The bottleneck height
decreases with increasing $R_{\lambda}$, but relatively high $R_{\lambda}$
are necessary to make the effect vanish completely. Based on nonlinear
and nonlocal shell-to-shell energy transfer Verma \& Donzis \cite{verma_energy_2007}
estimate that the bottleneck is basically absent for $R_{\lambda}>10^{4}$,
but acknowledge that this might be an overestimate. While lending
support to existing studies of the bottleneck effect, especially \cite{donzis_bottleneck_2010}
and theories that incorporate a $R_{\lambda}$-dependence of the peak
height, an investigation of the effect in terms of absolute measurements
of spectra seems necessary to confirm these claims experimentally.
With subminiature probes of low thermal lag, such a study might be
possible with a constant current anemometer, whose frequency response
is intriniscally more simple.

\section{Conclusions}

Hot-wire measurements of high-wavenumber parts of the energy spectrum
in a turbulent flow such as the bottleneck effect are distorted by
non-flat frequency responses typical for constant temperature anemometers.
When the experiment allows for a change in $R_{\lambda}$ through
the forcing mechanism alone, the $R_{\lambda}$-dependence at such
high wavenumbers can be investigated. We have used an active grid
to fix the frequency at which the bottleneck effect occurs. Thereby
the bottleneck effect was always subject to the same systematic errors.
By considering spectra relative to a reference spectrum, we found
a $R_{\lambda}$-scaling of the bottleneck effect.

We have found indications that the bottleneck effect gets weaker with
increasing $R_{\lambda}$. The data are very similar to the results
of DNS \cite{donzis_bottleneck_2010} and a theory \cite{verma_energy_2007}.
$R_{\lambda}$ exceeds 5000 in the VDTT with an autonomous active
grid, which is unprecedented in any wind tunnel. At $R_{\lambda}>3000$
our data supports a further decrease of the bottleneck height with
$R_{\lambda}$ as well as a constant or absent bottleneck.

\section*{Acknowledgements}

The operation of the experiment would be impossible without the help
and expertise of A. Kubitzek, A. Kopp, A. Renner, U. Schminke and
O. Kurre. The NSTAPs were generously provided by M. Hultmark and Y.
Fan. We thank P. K. Yeung and T. Ishihara for providing the numerical
data. We thank D. Lohse and P. Roche for useful comments.

\appendix
%dummy comment inserted by tex2lyx to ensure that this paragraph is not empty

\section{A brief description of the wind tunnel}
 
The VDTT consists of two 11.7 m long straight cylindrical tubes connected
by two elbows of center-line radius of 1.75 m. The tunnel was filled
with sulfur-hexaflouride (SF6) at pressures between 1.5 and 15 bar
for the measurements presented here.

The flow is propelled by a fan rotating at up to 24 Hz creating mean
flow speeds of up to 5.5 m/s. It passes the first elbow and enters
a heat exchanger, which removes any turbulent energy dissipated into
heat and thus keeps the temperature in the tunnel constant. The rectangular
cross-section of the heat exchanger is smoothly adapted to the tunnels
circular geometry by contractions. The vertical slots of the heat
exchanger are expected to destroy large-scales structure present in
the flow. After the heat exchanger, the flow passes an 80 cm long
expansion, which adapts it to the measurement section. While passing
this expansion the flow is stabilized and homogenized by three consecutive
meshes of ascending spacing. The flow enters a 9 m long measurement
section through an 104 cm high active grid, which is directly followed
by a 70 cm long expansion to the measurement section's height of 117
cm. The measurement section is followed by another elbow and enters
a second measurement section through another sequence of three meshes
before being accelerated again by the fan.

\section{Data Acquisition and Analysis Procedure }

The NSTAPs were operated following largely \cite{fan_high_2017}
using a Dantec StreamLine 90C10 module within a 90N10 frame. The CTA
bridge was set to a 1:1 ratio and the overheat is determined by an
external resistor $R_{\text{ext}}$ connected to the system. Typical
overheat ratios $R_{\text{ext}}/R_{\text{Probe}}$ were 1.2-1.3, where
$R_{\text{Probe}}\sim100\,\Omega$ denotes the probe cold resistance.
The Dantec wires were used in a 1:20 bridge utilizing the internal
automatics to set the overheat. The data was acquired in the following
procedure: The hot wire frequency response and proper operation was
tested on a very basic level using the square wave test built into
the Dantec CTA-system. The hot-wire system was calibrated by scanning
a range of mean flow speeds set by the fan frequency in the tunnel.
We determined the mean flow speed through the differential pressure
between a pitot tube and a static pressure probe. The differential
pressure was picked up by a Siemens SITRANS differential pressure
transfucer/ We chose \textasciitilde{} 20 calibration points spaced
by \textasciitilde{} 0.1 m/s. The probe voltage was recorded for 60
seconds along with the mean pressure difference, a voltage-velocity
curve was calculated, and King's law was fitted to the data. In between
calibration points we waited for 45 seconds for the mean flow to become
stationary. The data was recorded with a National Instruments NI PCI-6123
16-bit DAQ-Card at sampling rates of 60 or 200 kHz. Higher sampling
rates were used for NSTAP measurements, where the CTA analog low-pass
filter was set to 100 kHz. When using standard hot wires, the filter
frequency was set to 30 kHz and the data was sampled at 60 kHz. The
data was recorded in segments of 6 million voltage samples, each saved
to disk in a 16-bit binary format.

We shall briefly outline the initial data analysis procedure used
to obtain essential turbulence statistics as well as the power spectrum.
Each of the following steps was carried out on each segment and the
results were averaged over all files in the end. We used King's law
with parameters obtained form the calibration data to convert the
voltages to velocities. Note that the shape of the energy-frequency
spectrum is independent of the calibration, which is only required
to obtain its absolute value. Because the analog filtering was not
sufficient to filter out all noise, we low-pass filtered the data
digitally using a sinc-Filter in forward and reverse directions. This
introduces edge effects, which we remove by cutting the first and
last 60 points of the time series. We then subtract the mean $U$
from the velocity time series to obtain a time series of $u$. The
remaining analysis is performed on this filtered dataset. The power
spectra were calculated using MATLB's fft function, which is based
on the FFTW-package . We calculate the correlation function using
MATLAB's xcorr function, which itself relies on the aforementioned
fourier transform procedure as well as structure functions of order
1 to 8. Finally, we obtain histograms of velocity and voltage. We
use Taylor's Hypothesis, which assumes that a one-dimensional velocity
field can be obtained from a time series by multiplying the time increments
by the mean velocity: $\Delta x=\Delta t\cdot U$. The power spectra
are normalized using the assumption that $\int E(k)dk=u^{2}$.

We routinely calculate basic turbulence quantities in different ways
and check the results for consistency. The quantites $R_{\lambda}$,
and $\eta$ depend on the mean energy dissipation rate $\varepsilon$,
which we measure using the third-order structure function $S_{3}(r)=\langle(u(x+r)-u(x))^{3}\rangle=4/5(\varepsilon r)$.
The last step follows from the Navier-Stokes equations and is also
predicted by Kolmogorov's 1941 theory. In practice we estimate $\varepsilon=\max(5/4\ S_{3}/r)$
and check the result with $\varepsilon=15\nu\int k^{2}E(k)dk$, and
$\varepsilon=\max(S_{2}^{3/2}/r)$. The integral length scale is calculated
as $L=\int_{0}^{\infty}C(r)dr$, where $C(r)=\langle u(x+r)u(x)\rangle$
is the velocity auto-correlation function. Its error mainly stems
from the ambiguous choice of the upper integration limit, which leads
to a relative error about 10\% in $L$.

\begin{figure}
\includegraphics[width=1\columnwidth]{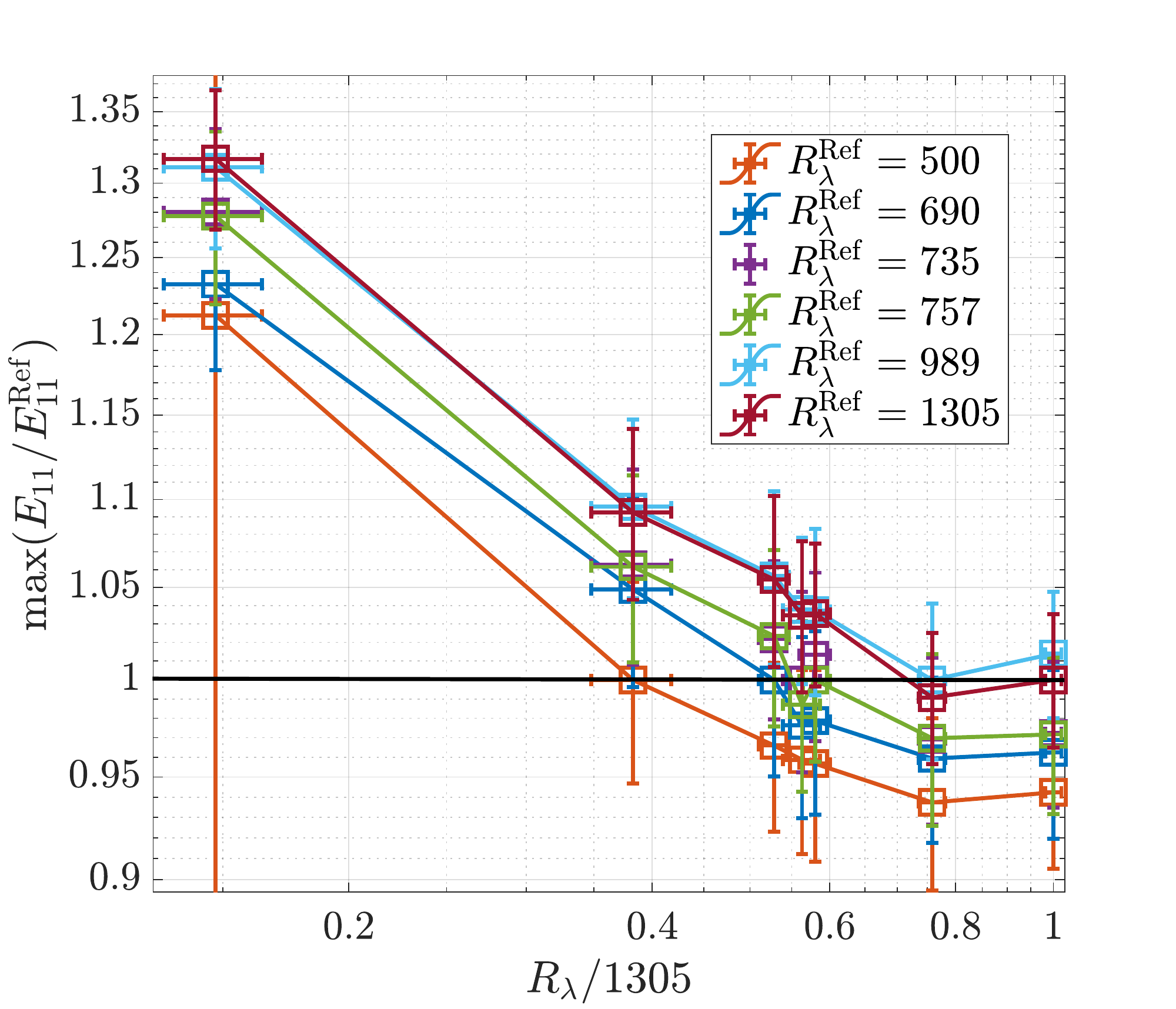}

\caption{The bottleneck height in Dataset 1 with different choices of $E_{11}^{\text{Ref}}$.
If the analysis is independent of the choice of reference spectrum,
the graphs are parallel in this representation. Note that the reference
spectrum always has bottleneck height 1. \label{fig:IndependenceLow}}
\end{figure}

\begin{figure}
\includegraphics[width=1\columnwidth]{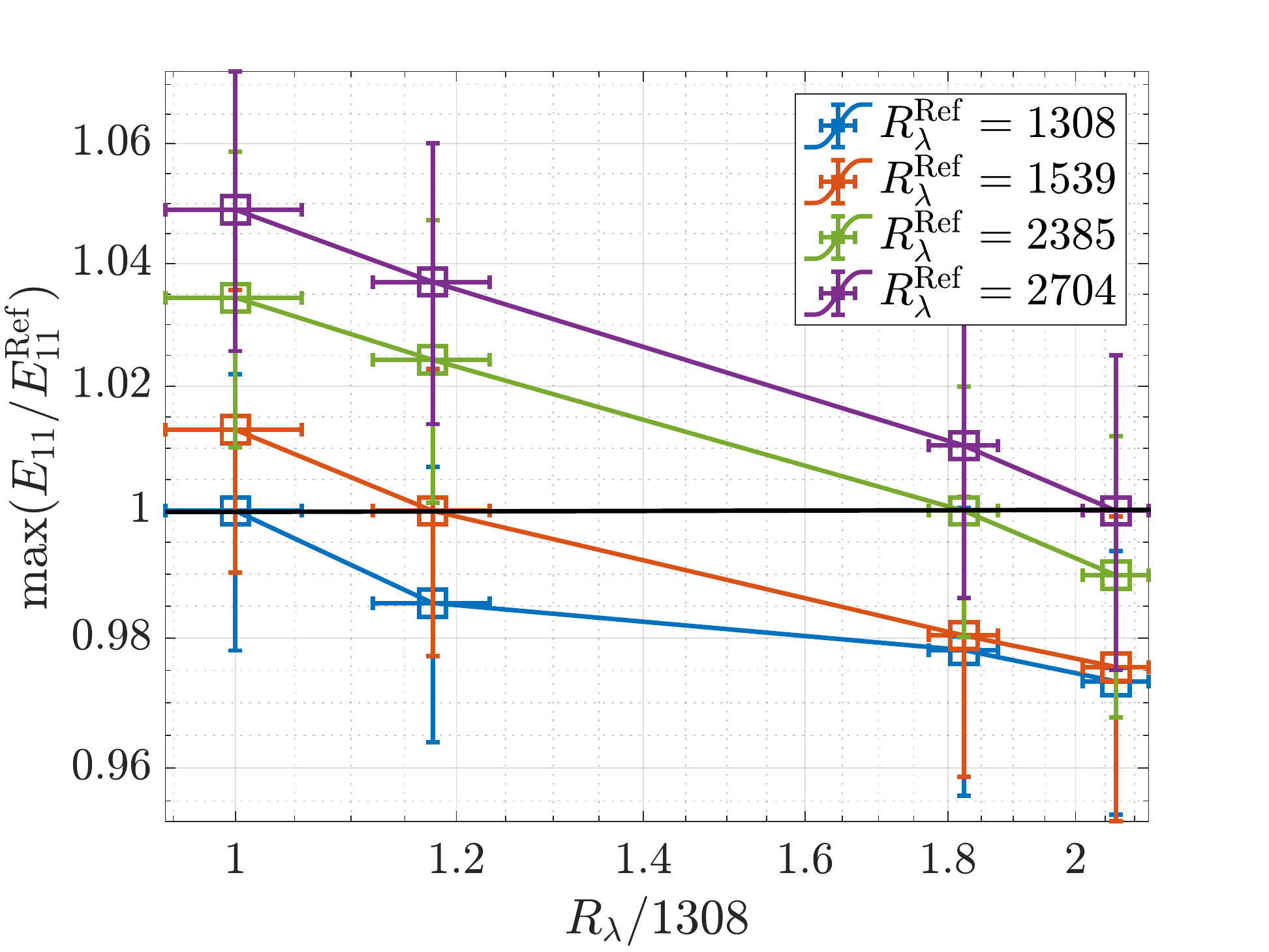}

\caption{Same as Fig. \ref{fig:IndependenceLow} for Dataset 2}
\end{figure}

\section{Calculation and Cross-check of Relative Spectra}

To obtain relative spectra, the initial spectrum consisting of 3 million
points was downsampled to 50 000 logarithmically spaced datapoints.
To remove the noise from these spectra, we have smoothed them using
a fractional octave smoothing algorithm. It multiplies the spectrum
at each frequency with a Gaussian centered around the current frequency
$f_{i}$ with a width of $\sigma_{i}=(f_{i}/n)/\pi$, where $n$ determines
the smoothing level. Therefore, the smoothing window is larger for
higher frequencies. To estimate the noise level in the spectrum and
the associated statistical error, we consider the data within $3\sigma_{i}$
of each frequency. We estimate the standard error as $\delta=3\sqrt{\text{Var}/N}$,
where $N$ is the number of points considered and $\text{Var}$ denotes
their variance. Finally, the compensated spectra are calculated as
$\psi(k)=E_{11}(k)k^{5/3}\varepsilon^{-2/3}$, which can be written
as $\psi(f)$ by Taylor's Hypothesis. Finally, we divide the $i$-th
spectrum in a dataset by the reference spectrum: $\psi_{i}(f)/\psi_{\text{Ref}}(f)$.
The result is normalized at $k\eta=0.015$ to remove offsets introduced
by uncertainties in $\varepsilon$ and to simplify comparisons.

\begin{figure}
\includegraphics[width=1\columnwidth]{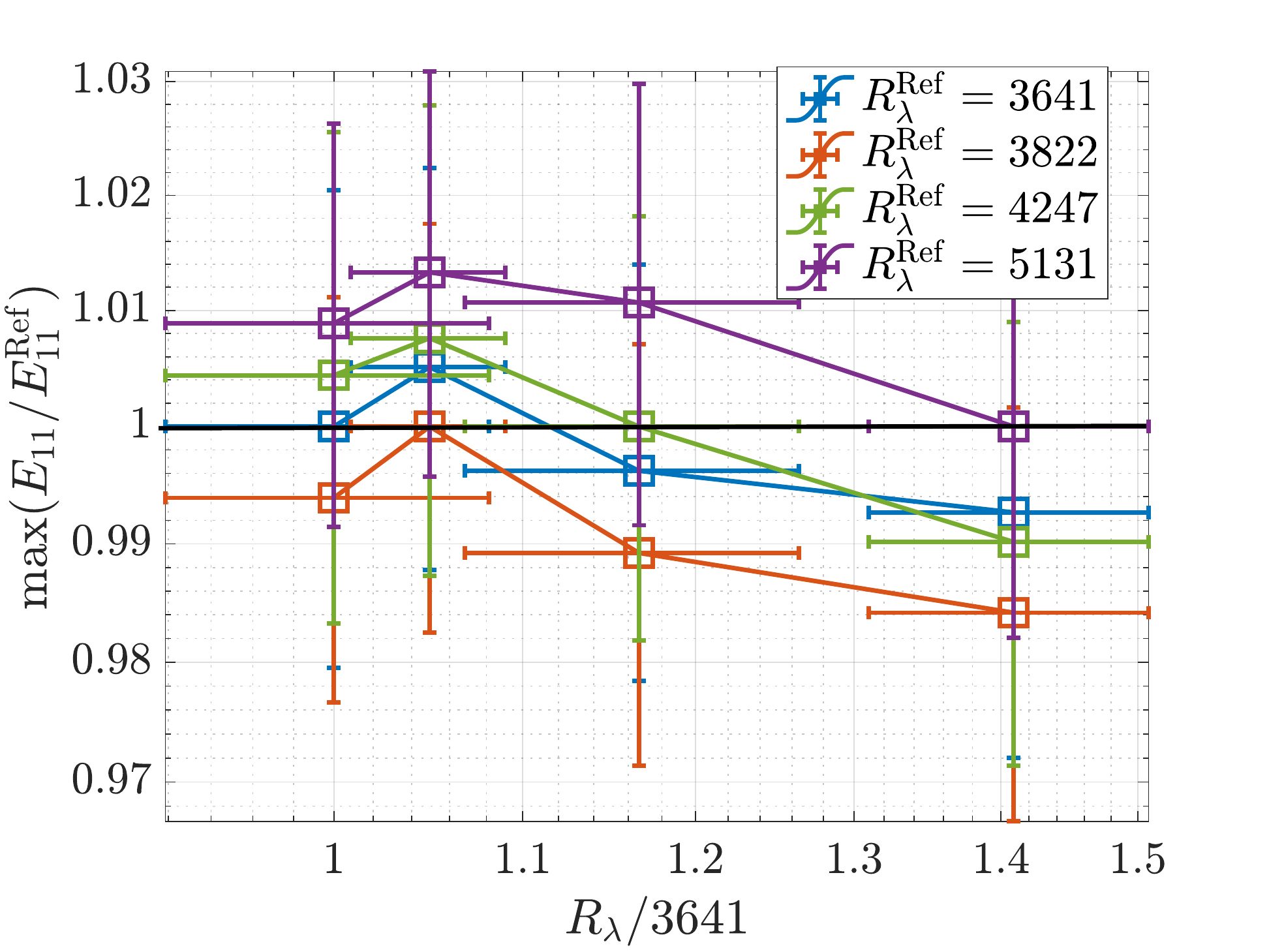}

\caption{Same as Fig. \ref{fig:IndependenceLow} for Dataset 3 \label{fig:IndependenceHigh}}
\end{figure}

An important cross-check of the technique is its independence from
the choice of reference spectrum. To this end we have calculated the
bottleneck effect according to the analysis outlined above for all
possible choices of reference spectra. The results are shown in Figs.
\ref{fig:IndependenceLow}-\ref{fig:IndependenceHigh}. They show
the peak height in the relative spectra as a function of $R_{\lambda}/R_{\lambda}^{\text{Ref}}$.
$R_{\lambda}/R_{\lambda}^{\text{Ref}}$ has been normalized to the
value that was chosen in the main part of the paper to increase the
clarity of the figures. If the analysis is independent of the choice
of reference spectrum $E_{11}^{\text{Ref}}$, a different choice $E_{11}'^{\text{Ref}}$,
should move the resulting curve by a factor of $E_{11}^{\text{Ref}}/E_{11}^{\text{Ref}}$
upwards and $R_{\lambda}^{\text{Ref}}/R_{\lambda}'^{\text{Ref}}$
to the right. The latter is trivial and has been removed from Figs.
\ref{fig:IndependenceLow}-\ref{fig:IndependenceHigh} by the additional
normalization. Thus, if the spectra are independent of the choice
of reference spectrum, the bottleneck curves should be parallel. Figs.
\ref{fig:IndependenceLow}-\ref{fig:IndependenceHigh} show that this
is valid in good approximation showing that the analysis is largely
independent of the choice of reference spectrum within a dataset.

\begin{figure}
\includegraphics[width=1\columnwidth]{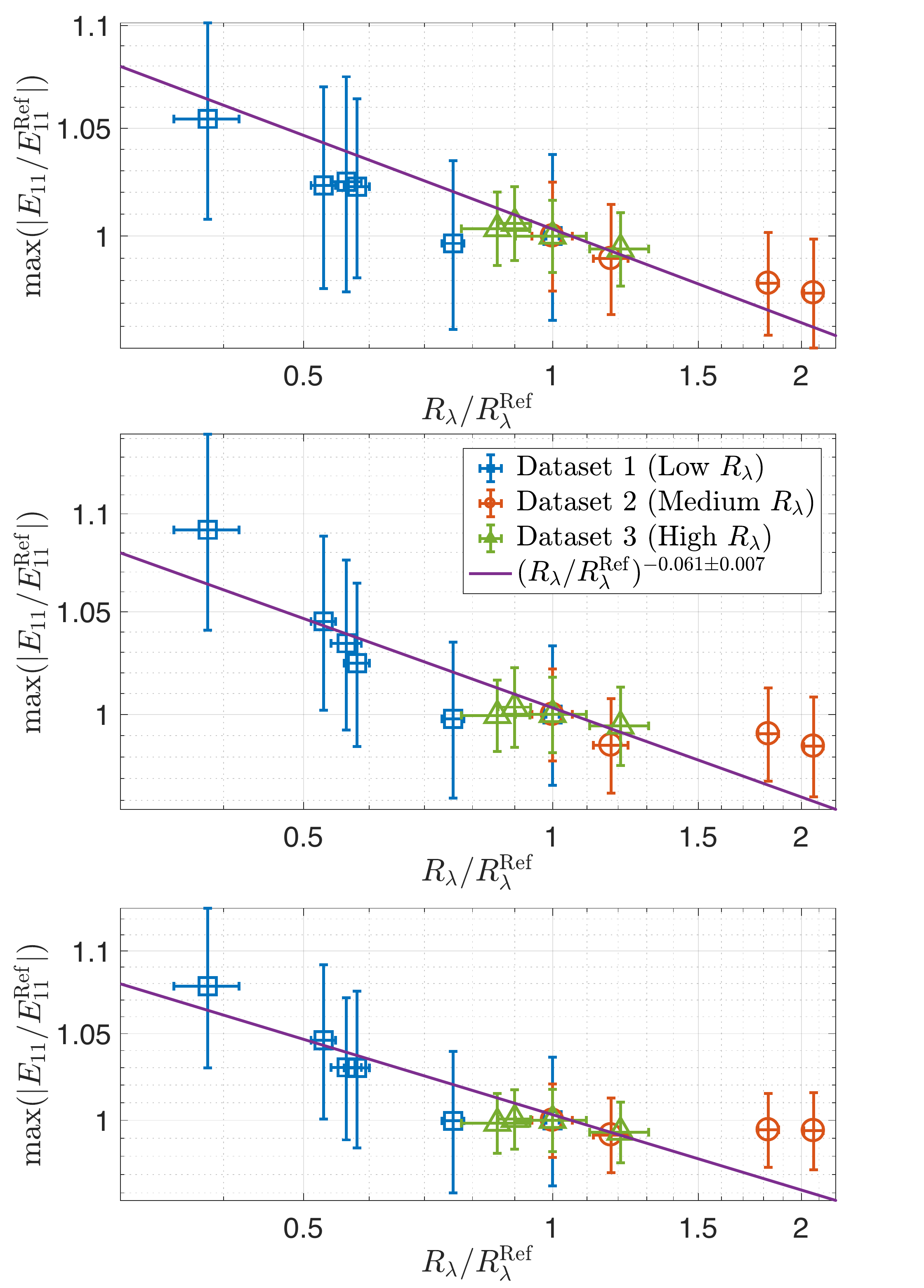}

\caption{Bottleneck height as function of $R_{\lambda}/R_{\lambda}^{\text{Ref}}$
with different bottleneck definitions based on the height of the relative
spectra at a fixed $k\eta$. The power law is the one found using
the initial definition of the bottleneck height in Fig. \ref{fig:Bottleneck-height-RelativeRe}.
Upper graph: Bottleneck at $k\eta=0.03$, middle graph at $k\eta=0.04$,
lower graph at $k\eta=0.046$ (as predicted by DNS). Datasets 1 and
2 still follow the trend found in the main part. Dataset 3 is not
inconsistent with the claim of a constant bottleneck height at $R_{\lambda}>3000$.
\label{fig:DifferentBNDefs}}
\end{figure}

\end{document}